\documentclass[prd,twocolumn,a4paper,showkeys,superscriptaddress]{revtex4}
\usepackage{graphicx}

\begin{document}

\newcommand{\be}{\begin{equation}}
\newcommand{\ee}{\end{equation}}
\newcommand{\bq}{\begin{eqnarray}}
\newcommand{\eq}{\end{eqnarray}}
\newcommand{\bsq}{\begin{subequations}}
\newcommand{\esq}{\end{subequations}}
\newcommand{\bc}{\begin{center}}
\newcommand{\ec}{\end{center}}
\newcommand{\al}{\alpha}

\title{Effect of Reconnection Probability on Cosmic   
       (Super)string Network Density}

\author{A. Avgoustidis}
\email[Electronic address: ]{A.Avgoustidis@damtp.cam.ac.uk}
\affiliation{Department of Applied Mathematics and Theoretical Physics,
Centre for Mathematical Sciences,\\ University of Cambridge,
Wilberforce Road, Cambridge CB3 0WA, United Kingdom}
\author{E.P.S. Shellard}
\email[Electronic address: ]{E.P.S.Shellard@damtp.cam.ac.uk}
\affiliation{Department of Applied Mathematics and Theoretical Physics,
Centre for Mathematical Sciences,\\ University of Cambridge,
Wilberforce Road, Cambridge CB3 0WA, United Kingdom}

\begin{abstract}
We perform numerical simulations of cosmic string evolution with
intercommuting probability $P$ in the range $5\times 10^{-3}\le P\le 1$, 
both in the matter and radiation eras, using a modified version of the  
Allen-Shellard code. We find that the dependence of the scaling  
density on $P$ is significantly different than the suggested   
$\rho\propto P^{-1}$ form. In particular, for probabilities   
greater than $P\simeq 0.1$, $\rho(1/P)$ is approximately flat, but  
for $P$ less than this value it is well-fitted by a power-law with
exponent $0.6^{+0.15}_{-0.12}$. This shows that the enhancement of  
string densities due to a small intercommuting probability is much  
less prominent than initially anticipated. We interpret the flat  
part of $\rho(1/P)$ in terms of multiple opportunities for string  
reconnections during one crossing time, due to small-scale wiggles.  
We also propose a two-scale model incorporating the key physical  
mechanisms, which satisfactorily fits our results over the whole  
range of $P$ covered by the simulations. 
\end{abstract}

\preprint{}
\maketitle

\section{\label{intro}Introduction}

 Cosmic strings have recently been reincarnated in the form of `cosmic
 superstrings' or, more specifically, fundamental strings and  
 one-dimensional $D$-branes produced at the end of brane inflation 
 \cite{DvalTye,Quevedo,SarTye,DvalVil} (for reviews see for example 
 \cite{PolchIntro,Kibblerev,DavKib,Majumdar}). These objects have 
 properties which can be significantly different than usual field   
 theory strings, and this opens up the possibility that cosmic string 
 observations could yield information about string theory, while  
 providing new ways to constrain various brane inflation models   
 \cite{Babichev,cycloops}. Cosmic superstrings typically have  
 smaller tensions, in the range $10^{-12} < G\mu < 10^{-7}$    
 \cite{SarTye,JoStoTye2,PolchIntro}, and reconnect with probabilities   
 which can be significantly less than unity \cite{JoStoTye2,PolchProb}, 
 unlike ordinary cosmic strings \cite{Shell_Recon}. This reduced   
 reconnection probability results in an enhancement of the string   
 number density today \cite{JoStoTye2,DvalVil,EDVOS}. Cosmic  
 superstring networks can also consist of more than one type of string,  
 which can zip together to produce trilinear vertices with links  
 stretching between them \cite{DvalVil,PolchStab}. 

 The evolution of cosmic superstring networks has recently been
 studied in Refs.~\cite{EDVOS,Sak,Mart_nonint,MTVOS,CopSaf}. The  
 key issue is whether these networks reach a \emph{scaling regime},  
 that is one in which their characteristic lengthscale stays  
 constant relative to the horizon. This self-similar evolution  
 is a well-known property of usual strings, which reconnect with  
 probability of order unity, at least in the abelian case where 
 no trilinear vertices can be formed (though some evidence for
 non-abelian scaling has also been found  
 \cite{VachVil,McGraw,McGraw1,MTVOS}). In all the above studies  
 of various models of cosmic supersting networks, evidence for   
 scaling behaviour was found, though the issue is not yet   
 completely resolved.  

 Assuming that an attractor scaling solution exists, an interesting   
 question to ask is how the density of the scaling network depends 
 on the intercommuting probability $P$. Using a simple one-scale   
 model \cite{Kibble}, Jones, Stoica and Tye have argued in  
 Ref.~\cite{JoStoTye2} that the expected behaviour is $\rho \propto  
 P^{-2}$. This is a dramatic effect, as a probability of $10^{-2}$, 
 for example, would lead to an enhancement of the string density by  
 a factor of $10^{4}$ compared to ordinary strings: we could
 live in a universe packed with $F/D$-string relics from a brane
 inflation era, which could in principle be observed to extract
 information about physics at the string scale! However, string  
 intercommuting is a small-scale process and can be expected to  
 depend crucially on small-scale wiggles on strings, which are  
 not captured by the simple one-scale model. One could still use  
 a one-scale (or VOS \cite{vos0,vosk}) model but introduce an
 \emph{effective} intercommuting probability $P_{\rm eff}=f(P)$ 
 \cite{EDVOS}, whose form should be determined by numerical simulations 
 or an analytic model for small-scale structure. One should expect that,   
 since small-scale wiggles would mean that strings may have more than  
 one opportunity to reconnect in each encounter, the effect of $P$
 on string density could be somewhat counterbalanced by the strings'  
 small-scale-structure. Thus a power law weaker than $\rho \propto  
 P^{-2}$ may be anticipated.  

 Indeed, Sakellariadou recently performed flat space simulations  
 of cosmic strings with reconnection probabilities in the range
 $10^{-3}\le P \le 1$ and found a power law $\rho \propto P^{-1}$  
 in the range $10^{-3}\le P \le 0.3$ \cite{Sak}. The purpose of this  
 letter is  to investigate the dependence of the string energy density
 $\rho$ on the intercommuting probability $P$ for strings evolving in
 more realistic cosmological backgrounds. We perform numerical
 simulations in both the matter and radiation era, and find a power   
 law significantly weaker than that of Ref.~\cite{Sak}. We explain our  
 results in terms of string small-scale structure and the importance  
 of a second scale in the problem. We thus propose a two-scale analytic 
 model which provides a good fit to our numerical results.

\section{\label{simul}Simulations}  

 We have performed numerical simulations of strings, with reduced
 intercommuting probabilities and evolving in FRW spacetime, using  
 a modified version of the Allen-Shellard code \cite{AllShel}. 
 The code sets up Vachaspati-Vilenkin initial conditions
 \cite{VachVil_IC} in a horizon volume within a fixed comoving box,  
 and evolves the system until the comoving horizon $\tau$ grows to  
 half the size of the box, when the evolution is stopped. The initial 
 networks had a resolution of 16 points per correlation length, and 
 constant resolution in \emph{physical} coordinates was enforced. Each 
 network was given a different intercommuting probability in the range
 $5\times 10^{-3} \le P \le 1$ and was evolved both in the matter and
 radiation eras, for a dynamical range ($\tau_{\rm final}/\tau_{\rm
 initial}$) of order 3, taking several days of CPU time.  

\begin{figure}
  \includegraphics[height=2.7in,width=3.0in]{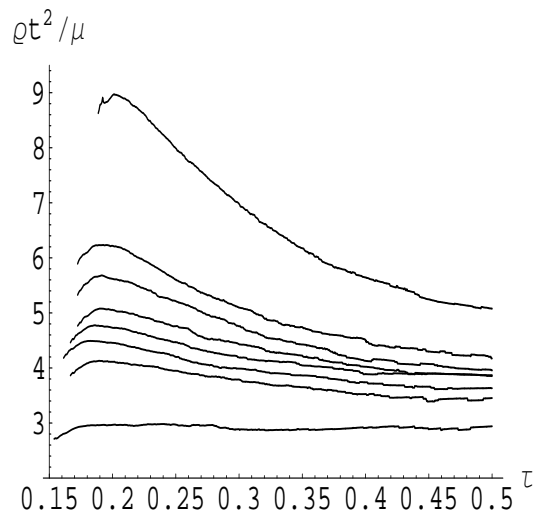}
  \caption{\label{bracket_fig} Dimensionless string density plotted  
           against conformal time $\tau$ for a network with
           intercommuting probability $P=0.75$. Each curve corresponds
           to a different initial horizon to correlation length
           ratio. The asymptotic curves bracket the scaling solution,
           which can be estimated to be $\rho t^2 / \mu=3.6^{+0.2}_{-0.1}$.}
 \end{figure}

 For each network characterised by a given intercommuting  
 probability $P$, approximately ten runs were  performed, each with  
 different initial string density (or precisely a different initial 
 horizon to correlation length ratio). By plotting the time evolution  
 of the string density for different initial conditions, one can  
 bracket the scaling solution, as shown in Fig.~\ref{bracket_fig}, 
 getting successively more accurate convergence with subsequent 
 runs. We have thus obtained, within errors, the scaling density   
 $\rho$ of these networks, each characterised by a given value of   
 $P$.  

\begin{figure}
  \includegraphics[height=2.7in,width=3.0in]{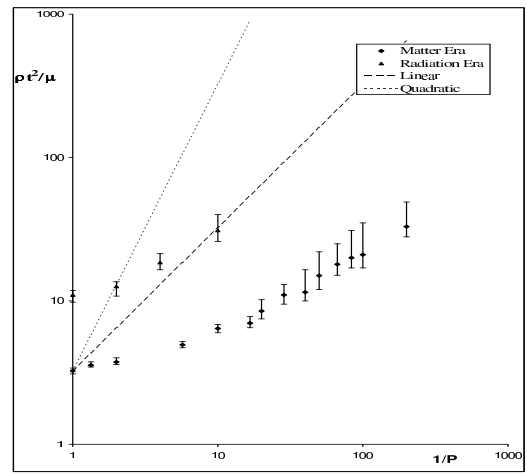}
  \caption{\label{rho_of_P_fig} Dimensionless string scaling density
           plotted against the inverse intercommuting probability
           $1/P$, for matter and radiation era runs. The constant
           slope part of the matter era data can be fitted by a power 
           law with exponent $0.6^{+0.15}_{-0.12}$. The overall dependence 
           of $\rho$ on $P$ is much weaker than the previously suggested  
           $\rho \propto 1/P^2$ and $\rho \propto 1/P^2$ forms.}
 \end{figure}

 In Fig.~(\ref{rho_of_P_fig}) we plot the dimensionless parameter   
 $\rho t^2/\mu$ (where $\mu$ is the string tension) versus the   
 inverse intercommuting probability $1/P$ for matter and radiation   
 era runs. In the former case $P$ ranges from $5\times 10^{-3}$ to   
 $1$, but in the latter our limited dynamical range has at present 
 only allowed us to bracket the scaling densities for $P$ in the   
 range $0.1\le P\le 1$. We see that for probabilities greater than   
 $P\simeq 0.1$, the function $\rho(1/P)$ is approximately flat, but   
 for smaller $P$ it develops a constant slope, on a log-log scale,   
 at least in the matter era. A weighted fit gives a slope of   
 $0.6^{+0.15}_{-0.12}$ for the matter runs, and the radiation era   
 data are consistent with this picture, though more data points are   
 needed to confirm the value of the slope. Comparing to the $\rho   
 \propto P^{-2}$ and $\rho\propto P^{-1}$ forms of Refs.~\cite{JoStoTye2}   
 and \cite{Sak} respectively (also plotted), we see that the   
 enhancement of string densities due to a reduced intercommuting   
 probability is much less prominent than what was initially anticipated.   
 For example, a probability of $5 \times 10^{-3}$ leads to an   
 enhancement in $\rho$ by only a factor of $10$, to be contrasted   
 with the predictions of $200$ (resp. $10^4$) obtained from $\rho   
 \propto P^{-1}$ (resp. $\rho \propto P^{-2}$).      
 
\begin{figure}[!h]
  \includegraphics[height=2.7in,width=3.0in]{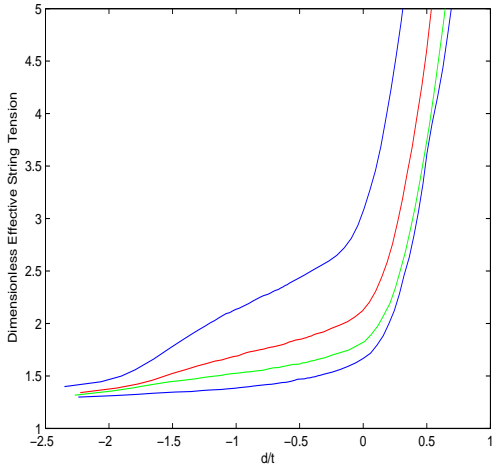}
  \caption{\label{mu_tilde_fig} Dimensionless effective string 
           tension $\tilde\mu$ plotted against physical distance  
           (measured in units of physical time t) for matter era  
           networks with $P=1,0.1,0.05$ and $0.01$. As $P$ is  
           reduced, the effective tension increases signalling   
           the accumulation of small-scale structure.}
  \end{figure}

 We have also investigated the effect of reducing the intercommuting
 probability $P$ on the amount of string small-scale-structure  
 \cite{MartShell_sss}. This can be quantified by the dimensionless  
 effective tension defined as $\tilde\mu=\mu_{\rm eff}/\mu$,   
 where $\mu_{\rm eff}$ is the effective energy per unit length at   
 the scale of the correlation length $\xi$ (this is the average   
 length beyond which string directions are not correlated). For   
 wiggly strings, $\tilde\mu$ is greater than unity, reflecting   
 the fact that string structure at sub-correlation length scales   
 `renormalises' the tension at the scale of the correlation length   
 to $\mu_{\rm eff}>\mu$ (refer to Martins \& EPS  
 \cite{MartShell_sss} for further discussion of these points). In   
 Fig.~\ref{mu_tilde_fig} we plot the dimensionless effective tension  
 $\tilde\mu$ against physical distance for a range of intercommuting  
 probabilities. Physical scales are expressed in units of the
 physical time $t$. As the intercommuting probability decreases,   
 we see a significant increase in the effective tension, especially  
 at the scale of the correlation length. This is strong evidence  
 for the accumulation of small-scale structure as $P$ is reduced.                 

\section{\label{discussion}Discussion}

 We can interpret the flat part of Fig.~\ref{rho_of_P_fig} in terms  
 of string small-scale-structure: if two colliding strings are wiggly,
 they may have more than one opportunity to reconnect during their  
 encounter.  On large scales, correlation length segments move
 together slowly and coherently ($v_{\rm c}\approx 0.15$ for $P=1$
 \cite{MartShell_sss}), but their small-scale structure is relativistic  
 and rapidly oscillates ($v_{\rm rms} \approx 1/\sqrt2$). One therefore  
 expects the strings to cross several times as they pass, thus the  
 overall process should be described by an \emph{effective} probability  
 of reconnection $P_{\rm eff}>P$. Reducing $P$ by a factor of a few,  
 for example, should not have a big impact on the string density  
 \cite{EDVOS}. Indeed, we can understand the flat part of  
 Fig.~\ref{rho_of_P_fig}, and in particular explain the position  
 where the slope starts to change, by using the following simple  
 model for string interactions.  

 Consider two colliding straight strings, described on small-scales by
 a monochromatic mode of amplitude $A$ and period $T$. If $v_{\rm c}$ is 
 their relative coherent speed, the duration of the collision is  
 approximately $\Delta t\approx 2A/v_{\rm c}$. Now on small scales   
 the strings are oscillating with period $T$, so the number  
 of reconnection opportunities during their crossing time is  
 $N\approx \Delta t / T = 2A / v_{\rm c} T$. But since on these scales 
 string modes are propagating with velocity $c\sim 1$, $T$ is equal   
 to the wavelength of the mode, and so $A/T$ is a measure of string  
 wiggliness. In particular, for large amplitudes we may write  
 $1+A/T \sim \tilde\mu$ from which we can infer that 
 \be\label{N_int} 
  N\sim \frac{2(\tilde\mu - 1)}{v_{\rm c}}.  
 \ee
 Interestingly, this simple estimate gives about the right value for 
 the position where the curve changes slope in Fig.~\ref{rho_of_P_fig}.  
 Indeed, from Fig.~\ref{mu_tilde_fig} we see that for $P>0.1$, $\tilde\mu$
 is in the range $1.6\le \tilde\mu\le 1.7$, while we know that string
 velocities at the scale of the correlation length are approximately  
 $v\approx 0.15$. This gives $N\sim10$ intercommuting opportunities. 
 One therefore expects that to see a significant effect on the scaling 
 string density, $P$ should be reduced by a factor greater than about  
 $10$. 

 Furthermore, we can estimate the effective probability by considering  
 the  $N$ reconnection opportunities, assuming each has an independent 
 probability $P$, that is 
 \be\label{P_eff} 
  P_{\rm eff}=1-(1-P)^N.
 \ee 
 For $N\simeq 10$, a probability as low as $0.2$ yields $P_{\rm
 eff}\simeq 1$, and significant deviations from unity only occur 
 for $P\simeq 0.1$ for which $P_{\rm eff}\simeq 0.65$, 
 as observed in Fig.~\ref{rho_of_P_fig}. In the limit $P\ll 1$,  
 Eq.~(\ref{P_eff}) gives $P_{\rm eff}\simeq NP\approx
 [2(\tilde\mu-1)/v_{\rm c}]P$ and so $P$ can only be enhanced by a
 factor of order $10$ (see Fig.~\ref{mu_tilde_fig}), which results
 in an effective reconnection probability that is still much less  
 than unity. 

 We now turn to explaining the constant slope part of  
 Fig.~\ref{rho_of_P_fig} for small probabilities. First recall  
 that a simple one-scale model with $P<1$ predicts $\rho\propto P^{-2}$  
 in direct contradiction with our numerical results. As discussed
 above, the introduction of an effective intercommuting probability  
 can dramatically change this for relatively large $P$ (say $P>0.1$),
 thus explaining the flat part of Fig.~\ref{rho_of_P_fig}. However, to
 accommodate the constant slope part for $P\ll 1$ one would need  
 $P_{\rm eff}\propto P^{\kappa}$, with $\kappa$ a constant less
 than $1/2$. At present there seems no motivation for such a $P_{\rm eff}$; 
 instead, Eq.~(\ref{P_eff}) suggests that $P_{\rm eff}$ is linear
 in $P$ for $P\ll 1$. Thus, it appears that no one-scale model can  
 fit our numerical results for $P\ll 1$: as the string intercommuting  
 probability decreases, the one-scale approximation becomes
 increasingly poor.  This is in agreement with the results of  
 Ref.~\cite{ShellAll_GHV} in which  long string intercommutings  
 were switched off in numerical simulations of evolving strings,  
 while small loop production was allowed. It was found that this 
 prevented the string density from scaling, that is, the inter-string 
 distance $L$ was no longer proportional to $t$ (the solution being  
 $\rho\propto L^{-2} \propto t^{-7/8}$).  Nevertheless, the  
 actual correlation length $\xi$ along the string did scale 
 at approximately the size of the horizon ($\xi\propto t$).  

 This is very similar to the situation we observe in our simulations.  
 Reducing the intercommuting probability leads to an increase  
 of the string density and therefore a decrease of the characteristic   
 length associated with it (the interstring distance $L$). The correlation  
 length $\xi$, however, is comparatively unaffected by this, and stays   
 at a scale of order the horizon (Fig.~\ref{mu_tilde_fig}). This can be   
 understood by considering the two distinct mechanisms for producing  
 loops: (i) self-intersections of the same string and (ii) collisions  
 between long strings. The former tends to chop off small loops,  
 straightening the strings out (thus affecting mainly the  
 correlation length $\xi$) while also controlling the amount of  
 small-scale structure $\tilde\mu$. Small loop production alone   
 is a significant energy loss mechanism, but it is not sufficient
 to ensure scaling \cite{ShellAll_GHV}. On the other hand, long  
 string reconnections have much more dramatic effects
 introducing large-scale `bends' in the strings which catalyse  
 the collapse of large regions of string and the formation of 
 many more loops.  In contrast, these energy losses are   
 sufficient to govern the interstring distance $L$ and cause scaling. 
  
 Small loop production, the first mechanism, is not greatly affected  
 by reducing the intercommuting probability: once the 
 string is sufficiently wiggly, left and right moving modes  
 which fail to interact due to a small $P$ will  
 keep propagating and meet more incoming modes, with which  
 they will eventually interact and form loops. It might take longer,  
 but such interactions are inevitable even for $P\ll 1$. The relevant   
 question is whether enough of these interactions can take place in  
 each Hubble time in order to straighten the strings out at the
 horizon scale, but this seems reasonable given the much shorter 
 timescale on which small loop production operates.  It appears to 
 be confirmed by Fig.~\ref{mu_tilde_fig}, where there is only a very 
 weak build-up of small scale structure (a factor of 2-3 in $\tilde \mu$) 
 while $P$ changes by over two orders of magnitude. On the other  
 hand, long string intercommuting, the second mechanism, depends  
 more crucially on $P$: two colliding strings have a given interval  
 of time in which to interact, so intercommuting is no longer  
 inevitable for relatively small $P<0.1$. A reduced $P$ necessarily  
 means less string interactions and less energy dumped from the long  
 string network in the form of loops. This leads to an increase in the  
 long string density and thus a significant decrease of the characteristic
 length relative to the correlation/horizon scale.
     
 To obtain an analytic model for such networks, it is therefore   
 necessary to introduce two scales: a characteristic length $L$ (roughly 
 the interstring distance) quantifying the energy density in strings,
 and a correlation length $\xi$, defined to be the distance beyond which
 string directions are not correlated. Indeed, similar models have 
 appeared in the literature (e.g. the three-scale model of 
 Ref.~\cite{AusCopKib}), but for normal $P=1$ cosmic strings  
 it has been established that these two scales are comparable  
 so the one-scale (velocity-dependent) approximation  
 \cite{vos0,vos,vosk}  can be surprisingly accurate  
 \cite{vostests}. Here, we develop a two-scale velocity-dependent  
 model which fits our numerical results, again surprisingly well   
 given its simplicity. 
      
 Consider a string network, characterised by tension $\mu$,
 correlation length $\xi$ and energy density $\rho$. We define 
 the characteristic lengthscale $L$ of the network by $\rho\equiv 
 \mu/L^2$ and note that the number of strings per correlation volume 
 $V=\xi^3$ is $N_{\xi}=\rho V/\mu\xi=\xi^2/L^2$. If $v$ is the typical 
 string velocity, each string intersects $N_{\xi}-1$ other strings in   
 time $\Delta t=\xi/v$, so we have $N_{\xi}^2-N_{\xi}$ intercommutings  
 per correlation volume per time $\Delta t$. Assuming that each 
 intercommuting produces a loop of length $\tilde c \xi$ (this can  
 be done formally by integrating an appropriate loop production  
 function over all relevant loop sizes, which introduces the loop  
 production parameter $\tilde c$ \cite{book}) the energy loss due 
 to the formation of loops can be written as 
 \be\label{delta_rho_loops_2s}  
  \left(\frac{\delta\rho}{\delta t}\right)_{\rm 2-scale} =   
  \frac{(N_{\xi}^2-N_{\xi})v}{\xi^4}\mu\tilde c \xi =   
  \tilde c \rho \left(\frac{\xi}{L^2}-\frac{1}{\xi}\right).  
 \ee
 This is to be contrasted with the corresponding result for the  
 one-scale model 
 \be\label{delta_rho_loops_1s}  
  \left(\frac{\delta\rho}{\delta t}\right)_{\rm 1-scale} =   
  \tilde c \frac{\rho}{L}.  
 \ee    
 As an interesting aside, we note that the last term in  
 Eq.~(\ref{delta_rho_loops_2s}) $\propto 1/\xi$ has the same form   
 (though opposite sign) as the term which would need to be introduced  
 to account for direct small loop production (or string radiation); it  
 cannot itself cause $L$ to scale as $t$.  
     
 Our velocity dependent two-scale model can therefore be constructed by 
 using the usual VOS model equations \cite{vosk}, derived by performing  
 a statistical averaging procedure on the Nambu-Goto equations of motion  
 and energy momentum tensor, but using the phenomenological loop 
 production term (\ref{delta_rho_loops_2s}) instead of 
 (\ref{delta_rho_loops_1s}). The result is a system of two coupled  
 ODEs, governing the time evolution of the characteristic length $L$   
 and the average velocity $v$ of string segments:  
 \begin{eqnarray}\label{Ldt_2s}
   2\frac{dL}{dt} & =& 2\frac{\dot a}{a}L(1+v^2) + \tilde c v
   \left(\frac{\xi}{L}-\frac{L}{\xi}\right) \\
   \label{vdt_2s}
   \frac{dv}{dt} &=& (1-v^2)\left(\frac{k}{\xi}-2\frac{\dot a}{a}v\right)
 \end{eqnarray}      
 In Eq.~(\ref{vdt_2s}), $k$ is the so-called momentum parameter, which  
 is a measure of the angle between the curvature vector and the  
 velocity of string segments and thus is related to the smoothness  
 of strings \cite{vos0}. Note that we have made no attempt to derive an
 evolution equation for the correlation length $\xi$ as the
 simulations show that this depends weakly on $P$ and moreover, 
 unlike $L$, it remains comparable to the horizon (Fig.~\ref{mu_tilde_fig}).  
 We will neglect this weak dependence of $\xi$ on $P$ and take $\xi=t$  
 in Eqs.~(\ref{Ldt_2s}-\ref{vdt_2s}).   
 
 To apply the model (\ref{Ldt_2s}-\ref{vdt_2s}) to our simulated   
 networks, we have also introduced an effective intercommuting   
 probability given by (\ref{P_eff}), that is, we have replaced 
 $\tilde c$ in Eq.~(\ref{Ldt_2s}) by $P_{\rm eff}\tilde c$. As a 
 first approximation, we have neglected the dependence of $N$ on  
 $P$ (this can be understood in terms of the dependence of $\tilde\mu$
 on $P$, see Eq.~(\ref{N_int}) and Fig.~\ref{mu_tilde_fig}) and we have 
 simply taken $N=10$ (a rough average value over the range of $P$). 
 For the loop production parameter, we have used the value  
 $\tilde c=0.23$ of Refs.~\cite{vostests,vosk} which fits both  
 radiation and matter era runs in the $P=1$ case. 
 
\begin{figure}[!h]
  \includegraphics[height=2.7in,width=3.0in]{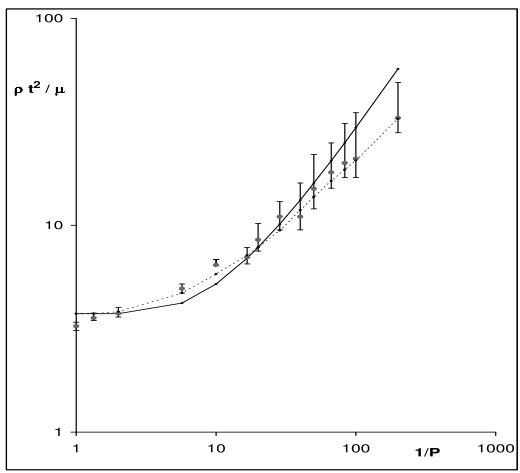}
  \caption{\label{2scale_fig} Scaling string density obtained 
           from simulations (data points with errors) and from the analytic 
           two-scale model (\ref{Ldt_2s}-\ref{vdt_2s}) (solid line). The  
           fit is improved further (dashed line) by phenomenologically 
           incorporating the dependence of the effective string tension 
           $\tilde \mu$ on the reconnection probability $P$.}
 \end{figure}

 In Fig.~\ref{2scale_fig} we plot the matter era scaling  
 energy densities, obtained both from the simulations and  
 our two-scale model (solid line), for string networks with  
 intercommuting probabilities in the range $5\times 10^{-3}\le   
 P\le 1$. We see that the model provides a surprisingly good  
 fit  (given its simplicity and the approximations made) of  
 the numerical data and it reproduces the observed change of 
 slope around $P\approx 0.1$. We stress that we have made no  
 special parameter choices to obtain this fit, and by modifying  
 parameters further we could do much better.  
 
 Nevertheless, it is clear from the dashed line in Fig.~\ref{2scale_fig}  
 how the model could be improved by taking into account the dependence  
 of the collision number $N$ on $\tilde\mu(P)$.  In this case, we have 
 estimated the effective string tension $\tilde\mu$ as a function of $P$  
 from Fig.~\ref{mu_tilde_fig}, which yields $N$ from Eq.~(\ref{N_int})  
 (that is, 6--18 over the full range of $P$) and then $P_{\rm eff}(P)$  
 from Eq.~(\ref{P_eff}) which is finally input `by hand' in our two-scale  
 model. The striking agreement that results provides strong motivation  
 for understanding the key physical mechanisms governing the small-scale  
 structure parameter $\tilde\mu$. Indeed, modifying $P$ seems to provide  
 a useful testbed for observing the dynamics of $\tilde\mu$, but we  
 leave a more sophisticated analysis for a future publication  
 \cite{AMS_inprep}. We note the need also to take into account the  
 dependence of $\xi$ on the reconnection probability, and a direct  
 small-loop production term (as discussed above). However, here, we  
 just highlight the fact that a simple version of our two-scale model  
 with the addition of the effective probability of Eq.~(\ref{P_eff})  
 seems to provide a satisfactory fit to the numerical data.

\section{\label{conc}Conclusion}   

 We have performed numerical simulations of cosmic strings with
 intercommuting probabilities in the range $5\times 10^{-3}\le P 
 \le 1$ evolving in a matter- or radiation-dominated FRW universe. We  
 have found that the dependence of the string density $\rho$ on  
 the intercommuting probability $P$ is much weaker than previously  
 suggested in the literature. In particular, the function $\rho(1/P)$  
 has an initially flat dependence, up to probabilities of about
 $0.1$, and then develops a constant slope (on a log-log scale) of  
 about $0.6^{+0.15}_{-0.12}$. This yields very different predictions 
 from the generally expected $\rho \propto P^{-1}$ form; a probability  
 of $P=5\times 10^{-3}$, for example, gives an energy density  
 enhancement of a factor of $10$, compared to a factor of $200$  
 with $\rho \propto P^{-1}$. Clearly, the results presented here are 
 important for determining the quantitative observational predictions  
 of models with cosmic (super-)strings. The distinction between  
 ordinary cosmic strings and strings with $P<1$ is obviously more  
 subtle than first anticipated. In particular, we note that cosmic  
 (super-)strings will not be as strongly constrained observationally  
 and many limits in the literature will have to be re-examined in  
 light of these results (or, at least, re-normalised). 
  
 We have also endeavoured to provide some physical explanations
 for why the string density depends non-trivially on the reconnection 
 probability. We can explain the flat dependence of $\rho(1/P)$ in 
 terms of small-scale wiggles on strings, which lead to multiple  
 opportunities for reconnection in each crossing time, thus introducing  
 an effective reconnection probability $P_{\rm eff}=f(P)$. By  
 approximating the small-scale structure on long strings by a  
 monochromatic mode, we suggested a physically motivated form  
 for $P_{\rm eff}$, which adequately explains the large $P$  
 behaviour of $\rho(1/P)$. 
 
 The simulations demonstrated that, although the string density  
 increases for small $P$ leading to a reduction of the characteristic
 length scale (or interstring distance $L$) of the network, the
 correlation length $\xi$ (the length beyond which string directions  
 are not correlated) scales at a size comparable to the horizon. We  
 have explained how one can understand this fact in terms of the two 
 distinct mechanisms for loop production: the self-intersection 
 of wiggly strings which produces small loops and tends to straighten 
 the strings, determining the correlation length $\xi$, and long-string 
 collisions, which cause much greater loop energy losses, determining the 
 characteristic length $L$. We used a simple two-scale model to describe  
 the behaviour of $\rho(1/P)$ over the whole range of probabilities  
 probed by the simulations. Our model provides an adequate fit to  
 the numerical data, given its simplicity and the approximations  
 used. We leave a more detailed numerical investigation and further  
 improvements of the analytic model for a future publication.

\begin{acknowledgments}
The authors would like to thank Carlos Martins for many
discussions. A.A. is supported by EPSRC, the Cambridge European Trust
and the Cambridge Newton Trust. This work is also supported by PPARC
grant PP/C501676/1. 
The code used was developed by EPSS and Bruce Allen. The simulations
were performed on COSMOS supercomputer, the Altix3700 owned by the UK
Computational Cosmology Consortium, and supported by SGI, Intel, HEFCE
and PPARC.   
\end{acknowledgments}

\bibliography{InterProb}

\end{document}